\documentclass[epj]{svjour}
\usepackage{graphics}
\DeclareGraphicsExtensions{ps,eps,epsf}
\begin{document}
\title{Ageing of Natural Rubber under Stress}
\author{S. M. Clarke, F. Elias and E. M. Terentjev}
\institute{Cavendish Laboratory, University of Cambridge,
Madingley Road, Cambridge CB3 0HE, U.K.}
\date{\today}

\abstract{We report a dynamical-mechanical study of stress
relaxation at small deformation in a natural (polyisoprene) rubber
well above its glass transition temperature $T_{\rm g}$.  We find
that an almost complete relaxation of stress takes place over very
long relaxation periods, even though the elastic network is
retained. The relaxation rate and the long-time equilibrium
modulus are sensitive functions of temperature which do not follow
time-temperature superposition. Many characteristic features of
non-ergodic ageing response are apparent at both short and very
long times. We interpret the observed behaviour in terms of the
nature of rubber crosslinks, capable of isomerisation under
stress, and relate the results to recent models of slow glassy
rheology.
} 
\maketitle

\section{Introduction}
 Physical ageing, defined as a dependence of linear
response functions on the time passed since the system was
prepared, is a characteristic feature of non-equilibrium glassy
dynamics and a subject of much current attention from all areas of
theoretical and experimental condensed matter physics. Polymers
and rubbers are also expected to fall into this class of
non-equilibrium fluctuation-dissipation behaviour when at
temperatures below their glass transition, $T_{\rm g}$.  However,
above $T_{\rm g}$, polymers and their networks are expected to be
capable of reaching equilibrium and not to exhibit such ageing
effects.

Uncrosslinked polymeric materials exhibit a complex behaviour on
deformation, being able to flow like a liquid, or responding
elastically, like a solid, depending upon the time scale of the
deformation relative to the dynamics of the molecules.  This
duality of mechanical response is called viscoelastically and is a
subject of a great number of classical studies \cite{doi,mcrum}.
In contrast, rubbers or elastomers are unable to flow because the
polymer molecules are permanently connected or crosslinked into a
percolating network. Elastomers are normally defined in terms of
their ability to support a stress indefinitely on extension and to
recover their original shape once the deforming influence has been
removed. This behaviour can be expressed in terms of a linear
rubber elastic modulus, $\mu$, the ratio of the applied stress to
the strain as the frequency of deformation tends to zero so that
$\mu = G'(\omega =0)$. In an ideal Gaussian network $\mu \simeq
n_{\rm s}kT$ with $n_{\rm s}$ the average crosslinking density
(or, strictly, the effective density including the effect of
topologically permanent entanglements in the network). At
sufficiently high frequency the mechanical response of rubbers is
not different from that of a corresponding polymer melt, being
dominated by dynamic entanglement effects.

Below their glass transition polymeric materials become brittle
and exhibit creep behaviour when a stress is applied and the
resulting strain measured.  In the original study of polymer
glasses, Struik \cite{struick} has demonstrated that polymer
glasses exhibit ageing behaviour where the time scale of response
is scaled by the waiting time, $t_w$, measured from the moment the
glass was quenched.  As expected, above the glass transition
polymers did not exhibit ageing behaving, in accordance with laws
of equilibrium thermodynamics, although, it was speculated that in
the regions surrounding the network cross-links or filler
particles, the glass transition temperature could be effectively
increased due to the locally higher polymer density and loss of
`free volume'.

Interest in the fundamental nature of the glass transition has
always  been high, since this represents one of the few major
outstanding challenges in condensed matter physics.   The
development of models of `soft glassy rheology' (SGR)
\cite{sgr1,sgr2} has renewed the interest in experimental study of
mechanical properties in other systems with quenched disorder. The
SGR model considers the long-time relaxation of mechanical
response in materials where there is the opportunity for random
yields and local stress relief when in the deformed state.  In an
application of the model \cite{sgr2}, the elements of the material
are capable of `hopping' to a new position allowing local stress
relief. This hopping is enhanced under the influence of a strain
that adds a directional bias, providing an aspect of flow.  On
making such a move the elements reconnect in increasingly deeper
potential wells, which makes the next hop increasingly difficult.
The direct application of such a mechanism are in the rheology of
foam or cellular solids. Such models have been demonstrated to
exhibit a `glass transition' with the corresponding breaking of
time temperature invariance, and ageing phenomena.

In the related systems of `spin glasses', where magnetic spins are
trapped in states with glassy orientational correlations, there
have been reports of a characteristic temperature dependence of
the non-ergodic relaxation behaviour \cite{bouchaud}.  This work,
and experiments \cite{vincent,alberici}, indicate that at each
temperature only specific modes of relaxation contribute to the
overall relaxation processes.

In this study we present experimental results on the long time
relaxation  behaviour of a sample of natural rubber.  The data
show, unexpectedly, that the stress response to an applied strain
can relax to virtually zero, if one waits long enough, even though
the short time linear rubber modulus is essentially unchanged.  We
also have investigated the temperature dependence of the
relaxation. The behaviour is interpreted in terms of the models of
glassy rheology and spin glasses. We identify the relevant
elastomer microstructure, in the form of isomerising sulphur
crosslinks, which allows local stress relaxation by `hops' arising
from directionally biased breaking and remaking of crosslinks. We
argue that such a `transient network' property makes many
elastomers non-ergodic even well above their glass transition.

\section{Experiment}
The step-strain relaxation measurements were made on a custom
built stretching device as described elsewhere \cite{we}. In this
device the samples, cut into thin strips approximately $10\times
30\times 0.5$~mm, were clamped at two ends, and were extended in a
step-strain fashion by ~5\%, measured with a micrometer. The force
on the sample was carefully measured as a function of time. The
nominal stress was calculated from the sample dimensions prior to
extension. The stress relaxation was followed over a period of
(typically) 40 days.  Some effort was made to carefully control
the temperature over such an extended period, with further
addition of a temperature controlled environment enclosing the
active temperature controller that maintained the samples at a
typical working temperature $\sim 80^{\rm o}$C, with deviations
not exceeding $0.5^{\rm o}$C. Polymer melts are well known to
exhibit faster relaxation at higher temperatures. An elevated
temperature of $80^{\rm o}$C allows us to probe longer effective
relaxation times than at room temperature and also avoids any
effects of strain induced crystallisation. This particular working
temperature was selected as being well above the glass transition
temperature but avoiding thermal decomposition. There was no
attempt to control humidity.

The stress values were obtained in arbitrary units and converted
to $\hbox{N/m}^2$ (Pa) by calibration with weights. Although the
stress-measuring device (Pioden Controls Ltd) was specially
temperature compensated, calibrations were performed over a range
of temperatures.  Data will be presented as engineering strain
($\lambda= L/L_0$, where $L_0$ is the unstretched length and $L$
the extended length) and nominal stress ($\sigma$, equal to the
measured force divided by the original cross-section of the
sample).

The samples of polyisoprene natural rubber were taken from a
commercial product. These samples were characterised by controlled
thermal decomposition and NMR, and by swelling in dichloromethane
with subsequent gel fraction measurements. The experimentally
determined gel fraction was 94\%. Natural rubbers are complex
mixtures of biological components which will include proteins and
other chemical impurities not directly attached to the
stress-supporting network. The loss of some of these elements
could easily account for this small weight loss during swelling.
Natural rubbers can be crosslinked in a number of ways but the
most common vulcanisation approach employs polysulphur crosslinks
formed by reacting the polymer, essentially poly(cis-isoprene),
with sulphur, zinc oxide and an accelerator at elevated
temperatures (e.g. $160^{\rm o}$C) \cite{morrison}. To ensure that
there were no artifacts from chemical decomposition or further
crosslinking at the elevated temperatures of our experiments, the
samples were pre-conditioned by long annealing at $150^{\rm o}$C
before the experiments. The gel fractions determined after such
pre-conditioning were unchanged, within experimental error. All
samples were taken from adjacent parts of the same rubber sheet
and fresh samples were used for each experiment to maintain a
controlled and similar thermal and mechanical history.

It was convenient in these studies to compare the behaviour of the
natural rubber crosslinked with sulphur with another elastomer
crosslinked with more stable carbon-carbon and carbon-oxygen
covalent bonds. In this respect we prepared a weakly
photocrosslinked polyacrylate consisting of steryl acrylate, 0.3\%
w/w crosslinks of ethane 1,2 diol, diacrylate, initiated with
Ergacure 184 (CIBA). These synthetic samples were sandwiched
between two glass slides $200 \, \mu$m apart irradiated for 1hr to
ensure complete reaction, and then removed and cut to shape for
mechanical experiments.

\section{Results}
\begin{figure} 
\centerline{
\resizebox{0.47\textwidth}{!}{\includegraphics{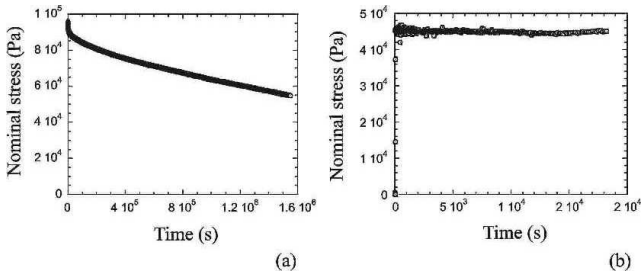}} }
\caption{(a) Stress relaxation in sample of natural rubber
extended by 5\%. Over the period of $\sim 18.5$~days the effective
linear modulus has fallen from $\sim 1.9$~MPa to $\sim 1$~MPa, by
almost a half of its value, and the relaxation is clearly
continuing. (b) Stress relaxation in polyacrylate elastomer is
unnoticeable after the first few seconds, the effective modulus is
constant at $\mu \sim 0.9$~MPa. } \label{fig1}
\end{figure}
Figure~\ref{fig1}(a) shows the typical stress relaxation for the
natural rubber after the imposition of a 5\% strain at $80^{\rm
o}$C. The most striking feature of this behaviour is that the
stress falls continuously over long periods.  In the subsequent
discussion we will show that the final stress value can fall to
essentially zero if one waits long enough.  Clearly, on very long
timescales this elastomer is not behaving elastically, but has a
viscous response. This behaviour should be compared with
Figure~\ref{fig1}(b) which presents data from the sample of
photocrosslinked polyacrylate under similar thermal and mechanical
conditions. These plots show that the stress in the synthetic
elastomer remains essentially constant (i.e. equilibrium) over
extended periods of time, where the natural rubber is relaxing
significantly.  This data also provides comfort that the
experimental device and environmental conditions are reasonably
stable.

A significant loss of elastic resistance in the natural rubber
illustrated in Figure~\ref{fig1}(a) could arise from the scission
of the polymer chains or cross-links of the network. The static
linear rubber elastic modulus $\mu$ would be an ideal indicator of
effective cross-link density $n_{\rm s}$ and the possible effects
of network degradation. It is impractical to wait infinitely long
for $G'(\omega \rightarrow 0)$ and, besides, the whole point of
this work is to show that additional physical but irreversible
processes take place in some networks. We, therefore, determine
the effective linear modulus $G'(\omega)$ after relatively short
times when we expect to probe the crosslink-dominated network
elasticity but not yet come into regime of significant transient
flow (see below). The choice of such a time (or, equivalently, the
effective frequency) for measurements of $G'(\omega)$ has to be
made carefully, taking into account the effects of chain
entanglements which will dominate at very short times. We
considered 1800s (0.5hr) relaxation sufficient to allow rapid
dynamic processes to have completed after each small step of
strain.

\begin{figure}[h]
\centerline{
\resizebox{0.47\textwidth}{!}{\includegraphics{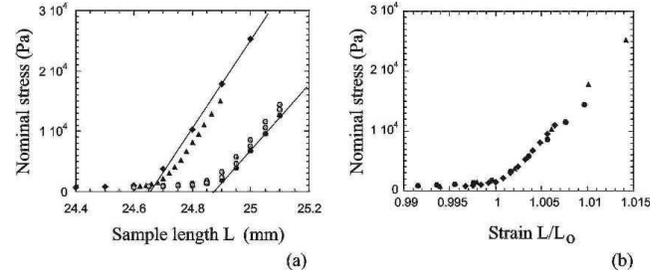}} }
\caption{ The short time stress strain response of the elastomer
as a means of extracting the elastic modulus.  (a) The
stress-extension behaviour in terms of absolute sample length,
measured before (filled diamonds) and after a 5\% extension with
the subsequent stress relaxations for 5 hrs (filled triangles) and
188hrs (filled circles). Open circle series show the partial
recovery of the natural length of the `188hrs'-old sample, at
intervals 18hrs, 36hrs and 119hrs after the extension was removed.
\ (b) Same data, with the $x$-axis in terms of the strain
$\lambda=L/L_0$. This rescaling takes into account the
irreversible (albeit small) change in natural length $L_0$,
allowing all the data to collapse on the same stress-strain line
indicating that the effective linear modulus $\mu$ is essentially
unchanged. } \label{fig2}
\end{figure}

To illustrate the elongation of the sample after a period of
stress relaxation we plot in Figure~\ref{fig2}(a) the length of
the sample in absolute units (mm).  The point at which the stress
begins to rise significantly is taken as the natural length of the
sample. Figure~\ref{fig2}(a) presents the short time
stress-extension curves for the rubber before and after a
step-strain experiment (constant 5\% extension) of different
duration (0, 5 hrs and 7.5 days). The natural length (the
zero-stress length) of the samples clearly increases after the
stress relaxation, however, the gradients of the curves are
unchanged. This is more clearly illustrated if the $x$-axis is
converted from the absolute length of the sample to the units of
strain $\lambda$, Figure~\ref{fig2}(b). The data are now found to
all collapse on the same stress-strain curve. The measured elastic
modulus before and after the long stress-relaxation experiment is
$\mu \sim 1.7$~MPa, essentially unchanged within the experimental
error. We conclude that although there has been significant stress
relaxation, as in Figure~\ref{fig1}(a) corresponding to extension
duration of 188hrs, the material has not degraded or significantly
altered its elastomeric network integrity. Clearly, there has been
a rearrangement of the network resulting in stress relief, but the
essential network topology and the number of crosslinks remains
constant.

Although the survival of the elastomeric network is unambiguously
confirmed by this argument, we also notice the lack of the
recovery of original natural length of the sample after prolonged
extensions. Complete recovery of the natural length could be
discerned in a reasonable time for short periods of stress
relaxation. For long periods, such as 188hrs shown by circles in
Figure~\ref{fig2}(a), partial recovery was evident but it was
clearly incomplete. The open circles show the behaviour on the
modulus measurement performed 18, 36 and 119hrs after the constant
strain step was removed. Clearly, unreasonably long time periods
would be required for any significant recovery to be observed and
all the evidence suggests that the original length will not be
fully recovered.

\subsection{Temperature Effects}
\begin{figure} 
\centerline{
\resizebox{0.3\textwidth}{!}{\includegraphics{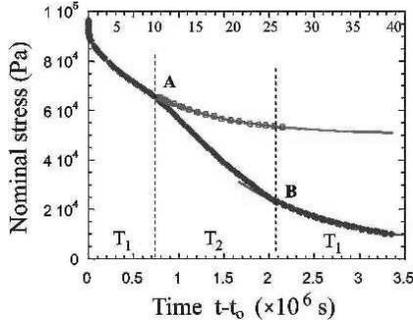}} }
\caption{ Stress relaxation in a sample of natural rubber after
imposition of a 5\% step-strain, with an intermediate temperature
change. The alternative time axis (figures on the top) gives the
number of days.  After a period of relaxation the temperature is
raised by $3^{\rm o}$ from $T_{\rm 1}=80^{\rm o}$C to $T_{\rm
2}=83^{\rm o}$C and then returned back to $T_{\rm 1}$. The solid
line is the extrapolation of exponential fit to the first region
of constant-$T$ data $\sigma_{\rm L}(t)$, eq.~(\ref{sigmaL}); the
circles over it are the shifted data points from the last region
where the temperature is returned to $T_{\rm 1}$ (see text).  }
\label{fig3}
\end{figure}

 As highlighted above \cite{bouchaud}, only particular modes
of relaxation are accessible at each temperature to magnetic spin
glasses.  On increasing the temperature, certain modes of
relaxation switch off and others switch on, with the reverse on
cooling.  We have imposed a temperature history to our sample to
make a comparison with this model.  Initially there is an extended
period ($\sim 12$ days) of stable relaxation at a temperature
$T_{\rm 1}$ (in this case $80^{\rm o}$C).  After this the
temperature was increased to a higher value $T_{\rm 2}$ (e.g.
$83^{\rm o}$C) and the relaxing stress data was collected for a
further 12 days. Finally the temperature is returned to the
initial temperature (e.g. $T_{\rm 1}=80^{\rm o}$C) and the final
stages of stress relaxation were monitored.  The effect of
temperature on the stress response is illustrated in
Figure~\ref{fig3}. The stress response during the initial period
at constant $T=T_{\rm 1}$, analogous to that in
Figure~\ref{fig1}(a), follows a simple exponential law of decay (a
variety of fast dynamic processes take place during the first few
minutes or even hours after the strain step; the simple
exponential decay is the longest apparent mode -- the relatively
short times are discussed in the next section). The corresponding
longest relaxation time $\tau_{\rm L} \sim 8.3 \times 10^{5}
\hbox{s} \approx 10$ days at $80^{\rm o}$C, the stress decay
represented by the extrapolated solid line in the
Figure~\ref{fig3}:
\begin{equation}
\sigma_{\rm L}(1) = 50430 + 38245 \, e^{-t/\tau_{\rm L}}
\label{sigmaL}
\end{equation}
(all units in Pa, the argument ``1'' signifying the first stage of
stress relaxation in Figure~\ref{fig3}) with the time $t$ measured
from the moment $t_{\rm o}$ when the extension has been applied.
This fit, when extrapolated to $t\rightarrow \infty$, implies that
the residual equilibrium value of stress $\sigma_{\rm eq}$ would
be $\sim 5 \times 10^4$Pa. The equilibrium rubber modulus would
then be $\mu \sim 1$~Mpa, almost a half of that observed during
the first hours of stress relaxation.

However, when the temperature increases, the relaxation rate is
seen to significantly increase and the extrapolated final stress
level decreases. In particular, it appears that the equilibrium
value of stress $\sigma_{\rm eq}$ that the system would have
reached remaining at constant temperature $T_{\rm 1}$, according
to eq.~(\ref{sigmaL}), has been surpassed already.

When, subsequently, the temperature returns to its original value
$T_{\rm 1}$ the relaxation resumes with the rate $\tau_{\rm L}$
that has been seen in the first stage. This portion of
experimental data is fitted by the exponential analogous to the
eq.~(\ref{sigmaL}) where we assume that the rate $\tau_{\rm L}$
and the pre-exponential coefficient take exactly the same values
as in (\ref{sigmaL}), but the time is now shifted by the amount
$\Delta$ -- the time rubber has spent in a different regime (at
$T_{\rm 2}$). In this way one obtains an excellent fit with only
one free parameter, the new value of the equilibrium stress level
$\sigma_{\rm eq}$:
\begin{equation}
\sigma_{\rm L}(3) = 7088 + 38245 \, e^{-[t-\Delta]/\tau_{\rm L}}
\label{sigmaL3}
\end{equation}
with $\Delta \approx 1.4 \times 10^6 \hbox{s}$ from the
Figure~\ref{fig3} and the same $\tau_{\rm L}\approx 8.3 \times
10^{5} \hbox{s}$ as in the eq.~(\ref{sigmaL}). As with the fit by
$\sigma_{\rm L}(1)$, we deliberately extrapolate the plot of
eq.~(\ref{sigmaL3}) outside the experimental data points, to
highlight the quality of the fit. Now, it appears, the rubber will
end up its relaxation with the effective linear modulus $\mu \sim
0.16$~Mpa, more than an order of magnitude lower than in the first
day under deformation.

Remarkably, the drop in the equilibrium stress (we obtain now
$\sigma_{\rm eq}(3) =7088 ~\hbox{Pa}$, instead of $\sigma_{\rm
eq}(1) = 50430 ~\hbox{Pa}$) is almost exactly equal to the
difference in experimental stress readings between the points A
and B in Figure~\ref{fig3}(a). In other words, the conclusion
appears to be that the modes of relaxation that were active during
the $T_{\rm 2}$-temperature interval are different and independent
of the relaxation modes at $T_{\rm 1}$. The amount of stress the
rubber has lost during the high-temperature interval, and the time
spent at $T_{\rm 2}$, are simply subtracted from the data. This
point is emphasized in Figure~\ref{fig3} by the following
transformation: we take the experimental data points in the third
region (the system returned to $T_{\rm 1}$) and shift them up in
$\sigma$ and back in time such that the point B is rejoined with
the point A. The experimental points fall onto the extrapolated
equation~(\ref{sigmaL}) for the whole available period of another
15 days, indicating that the physical processes in the initial
relaxation regime and that after the temperature step are exactly
the same, as if the temperature rise had not occurred.

Similar effects are observed on increasing the temperature from
$T_{\rm 1}=80^{\rm o}$C by $5^{\rm o}$ to $85^{\rm o}$C and by
$10^{\rm o}$ to $90^{\rm o}$C, except the fall in stress during
the period at the higher temperature is greater, as one might
expect. Again, the stress relaxation after the increment forms a
continuous curve with that before the increment. This data
indicates that the modes of relaxation $\sigma(t)$ and the final
equilibration value of $\sigma_{\rm eq}$ are sensitive functions
of temperature and also of the sample history, suggesting that the
traditional time temperature superposition principle will be
invalid here.

Our observation is similar to the behaviour of magnetic
susceptibility $\chi(t)$ in spin glasses \cite{vincent} and
dielectric susceptibility $\epsilon(t)$ in disordered dielectrics
\cite{alberici} (at constant applied strain, we too are measuring
the linear response function, the dynamic modulus $G(t)$~). In
these experiments the relaxation modes acting at one temperature
completely stop when the temperature is rapidly changed to a
different value and then re-continue when the temperature is
returned to the initial value.  The relaxation, as in our case,
forms a continuous line following the same time dependence as
before the temperature increment but just delayed by the time
spent at the higher temperature. This has been shown
\cite{bouchaud} to be an inherent feature of glass-like ageing
response.

\subsection{Relaxation at relatively short times}
\begin{figure} 
\centerline{
\resizebox{0.3\textwidth}{!}{\includegraphics{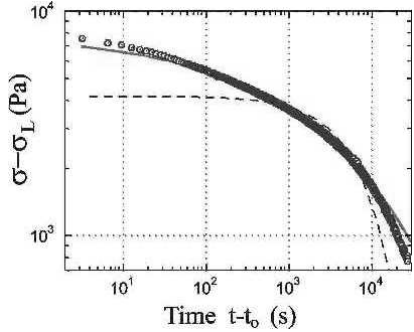}} }
\caption{The plot shows the relatively short time (first 7 hrs) of
stress relaxation. We plot the difference $\sigma-\sigma_{\rm L}$
to examine the fast initial relaxation processes not described by
the single-exponential behaviour $\sigma_{\rm L}(t)$. The solid
line gives the best fit, which turns out to be a stretched
exponential (see text). } \label{fig4}
\end{figure}

The analysis above has focused on the very long times of stress
relaxation, on the scale of many hours and days. The model
exponential decay $\sigma_{\rm L}(t)$, equation~(\ref{sigmaL}),
describes the data in Figure~\ref{fig3} only after a time of order
$8\times 10^4$s ($\sim 1$~day). At earlier times, which may be
considered `a long time' in a different sort of experiment (with a
corresponding frequency range $\omega \sim 10^{-5}$~Hz), the
deviations $\sigma(t) - \sigma_{\rm L}$ are substantial. This
deviation reflects the physical processes that take place in the
rubbery network if we neglect the longest-time effect of what we
expect to be a crosslink isomerisation under stress.
Figure~\ref{fig4} (plotted in the log-log format) shows the time
dependence during this relatively short initial time period.
The experimental data points $\sigma(t)$ with the subtracted model
decay $\sigma_{\rm L}$ are clearly faster than a power law, but
equally evidently slower than a simple exponential. An attempt to
fit the data by an exponential law is shown by the dashed line in
Figure~\ref{fig4}, to contrast with the best fit which is robustly
achieved by the stretched exponential with a power $\sim
\frac{1}{3}$, that is
\begin{equation}
\sigma(t)-\sigma_{\rm L} = 7400 \, \exp \left(-(t/\tau_{\rm
o})^{0.33} \right)    \label{short}
\end{equation}
(stress units in Pa) with relaxation time $\tau_{\rm o}
=3000~\hbox{s}$. We do not have an explanation for the observed
time dependence, modelled by the eq.~(\ref{short}). The exponent
$\sim 0.3$ of the stretched exponential relaxation has been known
to appear in a number of spin-glass problems, such as the
short-time relaxation of average magnetisation $m(t)$ after
field-quenching the spin glass \cite{bouchaud}.

In classical studies of rubber relaxation, e.g. of Ferry
\cite{ferry} and of Chasset and Thirion \cite{chasset}, the time
scale of experiments was usually under $10^5$s -- rather like our
`relatively short time regime' in Figure~\ref{fig4}. Authors
reported power-law dependencies of stress, $\sigma \sim
(t/\tau)^{-m}$, which has been attributed to the effect of
retracting of freely dangling chains on network deformation
\cite{curro}. We study the relaxation over at least an order of
magnitude longer time; it is also important to emphasize that
Figure~\ref{fig4} plots not the stress itself, but a difference
$\sigma(t)-\sigma_{\rm L}$ with the slowest mode subtracted.

\section{Discussion}
\begin{figure}[h]
\centerline{
\resizebox{0.47\textwidth}{!}{\includegraphics{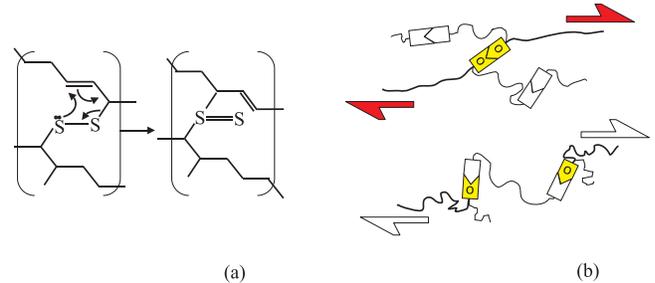}} }
\caption{Sulphur isomerisation mechanism (a) which allows the
reconnection of crosslinking bonds to a different location along
the chain. The scheme (b) shows how this mechanism of breaking and
re-connecting of linking points can lead to a local stress relief
between a given crosslink pair: the imposed deformation remains,
but is no longer resisted by an elastic force. } \label{fig5}
\end{figure}

 The significant loss of stress response on
extending a sample of natural rubber over a long period of time is
very surprising, particularly when combined with the fact that the
linear rubber modulus before and after such an experiment is
almost unchanged.  This behaviour is most likely linked to the
isomerisation of the sulphur-sulphur bonds that form the
crosslinks in this type of material.  Sulphur is the most common
means of vulcanising rubbers and its isomerisation in natural
rubbers has been reported before \cite{sulphur}.  An example of
the sulphur-sulphur isomerisation that could lead to the stress
relaxation but with retention of the crosslink is illustrated in
Figure~\ref{fig5}.  We note that the two chains are still
connected before and after the isomerisation but that the position
of the crosslink moves along the chain.  This isomerisation is
likely to be an activated process with some activation energy
$E_{\rm S}$, which should crudely determine the rate of relaxation
at a given temperature. Bias in the direction of crosslink
isomerisation is expected when the sample is under an external
strain, leading to local stress relief. This will continue until
there is insufficient external strain to allow further `hops' and
the `natural' barrier height will again control the kinetics.  In
this fashion the residual stress level $\sigma_{\rm eq}$ at long
time, in an relaxation experiment run at constant temperature, is
expected to be related to the barrier height.

The fact that the samples are able to recover their natural length
$L_0$ after relatively short periods under strain indicates that
the crosslinks are still in their original configuration in the
network and pull the material back to its original shape. As the
periods of stress relaxation become extended, more and more
isomerisation and stress relief events occur and the sample is
unlikely to ultimately return to its original shape. When,
eventually, the relaxing stress at fixed extension falls to its
low saturation value, $\sigma_{\rm eq}$, we expect that the sample
shape will not be recoverable at all. However, at any moment of
time, the total number of elastically active crosslinks in the
network remains the same and thus the effective rubber modulus
$\mu$ measured at different times remains the same too. The
measure of the time when the effect of crosslink isomerisation
becomes significant should be the same as that discussed in
Figure~\ref{fig4} for the decaying difference
$\sigma(t)-\sigma_{\rm L}$.

The fact that the (newly measured) linear rubber modulus $\mu$
recovers its value tells that the average number of effective
crosslinks in the system remains constant at any given time. This
sort of system has been examined theoretically, as the problem of
relaxation in a transient network where crosslinks (such as
hydrogen bonds) reorganise under stress. Flory \cite{flory} has
presented an intuitive model, later improved by Scanlan
\cite{scanlan}. Fricker and Edwards \cite{edwards,ed2} have
developed a statistical-mechanical formalism, with essentially the
same conclusion -- that, under a constant extension, the stress
remaining in the system where the crosslink breaking and
re-forming occurs at a rate $m^{-1}$ decays with time as
\begin{equation}
\sigma(t) \simeq \sigma_{\rm o} \exp \left( e^{-m \, t} -1 \right)
\label{flory}
\end{equation}
(the earliest and simplest result by Flory assuming the re-formed
crosslinks are not bearing stress and are thus permanent). We
could fit our data to this type of model relation equally well,
with the same number of (two) free parameters. We chose not to do
so because of an obvious discrepancy. According to these models, a
definite value of final stress $\sigma_{\rm eq}$ [approximately
37\% of the initial stress in eq.~(\ref{flory})] is predicted,
with temperature only affecting the rate $m^{-1}$. In contrast, we
find that at a given temperature there appear to be characteristic
modes of relaxation and long time stress relief, while different
modes become active at a different temperature, thus affecting the
final stress $\sigma_{\rm eq}$. Secondly, when the temperature was
returned to a lower value $T_{\rm 1}$ after a period at a higher
$T_{\rm 2}$, as in Figure~\ref{fig3}, the local rate of relaxation
(the tangent of $\sigma(t)$ curve) should be \underline{lower}
than that after the same time in a parallel experiment with no
temperature elevation period. Instead, we find unambiguously that
the relaxation rate at point B is \underline{higher}, in fact
equal to that at the earlier point A just before the temperature
rise, as is illustrated in Figure~\ref{fig3}. It appears that
different relaxation modes are active at each temperature, in
contrast to the thermally activated kinetic effects when {\it all}
modes are active simultaneously.

The behaviour we observe is rather analogous to the relaxation in
spin glasses \cite{bouchaud,vincent} where only specific modes are
accessible at each temperature. Recent models of soft glassy
rheology also highlight similar behaviour to that reported here.
The SGR model requires a microscopic mechanism of strain-enhanced
relief of stress and a distribution of well depths to exhibit a
whole range of glassy, non-ergodic behaviour. Our experimental
data clearly indicates that there is such a strain-induced relief
of stress. Additionally, because of the variety of sulphur-sulphur
links that will be present in the rubber, it is likely that a
broad distribution of potential well depths is also present. This
behaviour should be contrasted with an elastic network with
permanently fixed crosslinks, Figure~\ref{fig1}(b) where one finds
an essentially equilibrium response as is, in fact, expected in an
ideal elastic network above $T_{\rm g}$. In general terms, one
finds long relaxation and ergodicity breaking above $T_{\rm g}$
when the elastic degrees of freedom are coupled to an additional
randomly quenched field, which frequently arises in materials with
microstructure. In SGR, this corresponds to a model distribution
of yield events (as may occur in, e.g., a foam when cells
reconnect and move past each other \cite{sgr2}).  Here in our
study of natural rubber, we argue that the isomerisation of
sulphur crosslinks provides also provides such an effect and is
very close to the SGR model in its spirit.

The stress response fits well to an exponential time dependence
with the longest relaxation time of order 10-15 days.  This
contrasts with the earlier reported power-law behaviour, often
attributed to the effect of free chain retraction. Our results are
also in contrast with the reported logarithmic decay of linear
response function for the related process of creep \cite{derham}.
In the field of rubber technology the rate of relaxation is
usually expressed as the percent per decade \cite{pond} reflecting
this kind of time dependence. However, unlike the exponential
response we observe, such a logarithmic time dependence gives
unphysical behaviour at long times and so cannot provide physical
insight into the fundamental processes. This highlights the fact
that there are fundamental issues at the heart of rubber
technology left outstanding even after more than 50 years of
research. \\

\noindent The authors wish to thank Prof. Sir Geoffrey Allen,
Prof. Sir Sam Edwards, Dr K. Fuller and Dr P. Sollich for
interesting discussions.

\end{document}